# The physical limits of nanoLEDs and nanolasers for optical communications

Bruno Romeira and Andrea Fiore

*Abstract*—Nanoscale light sources are being intensively investigated for their potential to enable low-energy, high-density optical communication and sensing systems. Both nano-light-emitting diodes (nanoLEDs) and nanolasers have been considered, based on advanced nanophotonic concepts such as photonic crystals and plasmonic structures, with dimensions well into the sub-micrometer domain. With decreasing dimensions, light-matter interaction becomes stronger, potentially leading to efficient and ultrafast radiative emission, both in the spontaneous and stimulated regime. These features have created wide expectations for the practical prospects of such nanoscale light sources, in particular for optical interconnects. In this article we examine the limits to the downscaling of LEDs and lasers, and ask ourselves which type of source is most suited to ultralow-power optical communications. Based on simple physical considerations on the scaling of spontaneous and stimulated emission rates for semiconductor active regions at room temperature, we analyze the speed and energy limits for nanoLEDs and nanolasers as a function of their size. The role of spontaneous emission enhancement (Purcell effect) in practical nanophotonic sources is also revisited. The main conclusion is that nanoLEDs reach a fundamental energy/speed limit for data rates exceeding a few Gb/s, whereas nanolasers with active dimensions in the range of few 100s nm may enable direct modulation rates larger than 40 Gb/s at power levels adequate for short-distance and low-energy optical interconnects.

## I. Introduction

THE exponential increase of internet traffic sets demanding requirements on data communication technologies. Optical interconnects present higher bandwidth-distance products, lower electromagnetic interference and potentially lower power consumption than electrical interconnects [1], and are being deployed at increasingly shorter distances, for example within data centers. In the longer term, chip-to-chip and even intrachip communication may be performed with optics. However, traditional semiconductor lasers developed for long-distance optical communications operate very inefficiently at the low energy budgets characteristic of such short-distance links (~pJ/bit). As an example, an edge-emitting laser with a length of few hundred micrometers typically requires few tens of mW of electrical power just to reach threshold, corresponding to energies of few pJ/bit at data rates of 10 Gb/s. This is in fact many orders of magnitude larger than the optical energy required for photodetection (~20 photons or ~2.5 aJ/bit for an ideal, shot-noise limited receiver, and ~1000 photons or ~0.13 fJ/bit for a thermal-noise limited receiver [2]). The large threshold power is related to the need to achieve population inversion over a relatively large device area (~hundreds of $\mu m^2$). While a light-emitting diode (LED) does not present a threshold, conventional LEDs also operate inefficiently in macroscopic devices due to the small fraction of spontaneous emission which can be collected in the output channel. Additionally, their modulation bandwidth is typically limited by the spontaneous emission lifetime to <1 GHz. In analogy with the scaling of electronic circuits, the avenue for increasing the efficiency of optical sources at low energy/bit levels is clearly downscaling. Vertical-cavity surface-emitting lasers already present lower threshold currents due to the reduced active area, and they are widely deployed in low-power communication and sensing applications. Scaling to the submicrometer range requires more advanced optical confinement methods, such as 2D or 3D photonic crystals or the use of plasmonic resonances, and indeed such methods have been employed to fabricate nanoscale lasers and LEDs (see [3] for a recent review). An added benefit of the size reduction is that the rates of spontaneous and stimulated emission scale inversely with the mode volume, leading to faster and more efficient light emitters. In the case of LEDs, the increase of the spontaneous emission rate ("Purcell effect" [4]) has been suggested as a method to make emission into a given mode the dominant recombination process, and thereby improve the efficiency to the levels typical of lasers. In fact, due to the increased efficiency in the spontaneous emission regime, the characteristic laser threshold tends to disappear in nanolasers, so that the distinction between lasing and nonlasing devices is

This work was supported in part by Dutch Ministry of Education, Culture and Science under Gravitation programme "Research Centre for Integrated Nanophotonics". The work of Bruno Romeira was supported in part by the Marie Curie COFUND Programme - NanoTRAINforGrowth II and by the European Commission under the H2020-FET-OPEN project "ChipAI", grant agreement no. 828841.

B. Romeira is with the Ultrafast Bio- and Nanophotonics Group, INL – International Iberian Nanotechnology Laboratory, Av. Mestre José Veiga s/n, 4715-330 Braga, Portugal (e-mail: bruno.romeira@inl.int).
A. Fiore is with the Dep. Applied Physics and Institute for Photonic Integration, Eindhoven University of Technology, PO Box 513, 5600 MB Eindhoven, The Netherlands (email: a.fiore@tue.nl).

less evident than in macroscopic structures. Due to the increased emission rate, the modulation bandwidth of lasers and LEDs is also expected to increase with reduced size, potentially opening the way to modulation rates >40 Gb/s. In view of their potential for low-energy, high-speed optical interconnects, it is important to analyze the limits of scaling for practical nanoLED and nanolaser structures, and ask ourselves what kind of device (laser vs. LED) and what device size is most suited to efficiently provide the required optical energies/bit (~fJ) at the 10-100 Gb/s bit rates relevant for short-distance interconnects. In the following, after reviewing the basic physics of nanoscale light sources and some examples of their practical implementation in Section II, we apply a simple rate-equation model to study the scaling of nanoLEDs and nanolasers in Section III and IV, respectively. While disregarding many details of the device operation, our model captures the most important aspects of the scaling, namely the variation of spontaneous and stimulated emission rates, in a rigorous and self-consistent way, providing the ultimate limits for parameters such as modulation bandwidth and optical/electrical energies per bit. Assuming the material parameters of known optical semiconductors, we conclude that operation of a nanoLED at frequencies of 10 Gb/s and above is incompatible with the requirements of thermal-noise limited receivers. On the other hand, nanolasers with a mode volume in the 0.03 μm³ range could represent suitable sources for optical interconnects, with efficient operation and modulation bandwidths >40 GHz at fJ/bit energy levels, if optical losses, nonradiative recombination and device parasitics are kept at bay. Notably, we also conclude that one of the most important expectations for nanolasers — ultrafast modulation speed — is not obvious for deep-subwavelength lasers (<<0.03 μm³). As discussed in section IV, only at high, and probably unpractical, current densities >>100 kA/cm² one is able to achieve modulation speeds of tens of GHz.

## II. KEY INGREDIENTS

### A. Spontaneous and Stimulated Emission in Nanophotonic Structures

Essentially, nanoLEDs and nanolasers share a similar conceptual structure: An active material embedded in an optical cavity, Fig. 1(a). However, nanoLEDs are designed to operate in the spontaneous emission regime, implying that the number of photons in the cavity mode is lower than one and for this reason they are typically designed to have higher outcoupling rates. The cavity (or generally an optical structure designed to control spontaneous emission) is still needed to efficiently funnel spontaneous emission into the desired output channel. In contrast, nanolasers operate with lower cavity losses and photon numbers >>1, so that stimulated emission into the cavity mode becomes the dominant recombination process.

One of the key consequences of volume scaling in optical sources is the variation of radiative recombination rates. In order to properly understand and model both lasers and LEDs, it is crucial to remember that the spontaneous and stimulated emission rates into a given optical mode are related to each other through Einstein's relations, and that they both depend on the amplitude of the optical field at the emitter's position. In fact, the total emission rate into a given mode for a point-like emitter optimally coupled to the mode can be written as [5], [6]:

$$R_{cav} = \frac{\pi}{\hbar \varepsilon_0 \varepsilon_{ra}} \frac{d_{if}}{V}(N_{ph}+1) \int_0^\infty \rho(\omega) L(\omega) d\omega \qquad (1)$$

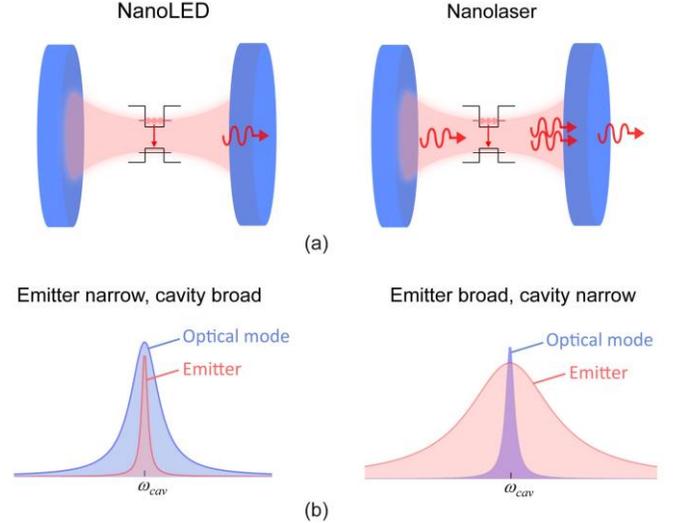

Fig. 1. NanoLED and nanolaser active materials embedded in subwavelength optical cavities. (a) Representation of a quantum emitter embedded in an optical cavity for the cases of a nanoLED (left) and a nanolaser (right). (b) Relation between the cavity resonance and the emitter's transition spectrum (assuming the emitter is resonant with the cavity mode ($\omega = \omega_{cav}$)) for (left) the "ideal" cavity quantum electrodynamic case (narrow emitter) and (right) the typical situation for a light source at room temperature (broad emitter).

Where $V$ is the mode volume, $d_{if}$ the matrix element of the dipole operator between the initial and final states, $\varepsilon_{ra}$ the relative dielectric constant in the active material, $N_{ph}$ the number of photons, $\rho(\omega)$ the density of optical states per unit of angular frequency and $L(\omega)$ the homogeneous broadening lineshape of the emitter. The contributions of stimulated and spontaneous emission are captured by the "$N_{ph}$" and "+1" terms and obviously scale in the same way. In particular, the inverse dependence on the volume comes from the fact that the field per photon scales as $\propto 1/\sqrt{V}$. This $1/V$ dependence is always present in the gain *and* spontaneous emission terms of laser rate equations, where it is often incorporated within the "confinement factor" [7]. Its effect on the spontaneous carrier recombination rate is however negligible in macroscopic lasers due to the large number of optical modes available to the emitter. This situation changes dramatically in the case of a small cavity, and can lead to the absence of a visible threshold in the input-output curves, as shown below.

Additionally, the emission rates depend on the relative alignment and width of the cavity spectral response $\rho(\omega)$ and emitter's lineshape, $L(\omega)$. Here a distinction between the "ideal" cavity quantum electrodynamic case (narrow emitter) and the typical situation for a semiconductor laser at room temperature (broad emitter) is important. If the emitter is much narrower than the cavity (Fig. 1(b) left), the integral in Eq. (1) simplifies to $\rho(\omega_{em})$ (where $\omega_{em}$ is the emitter's frequency), which at resonance is proportional to the inverse of cavity linewidth, $1/\Delta\omega_{cav}$. In other words, the emission (both spontaneous and stimulated) becomes faster for a cavity with lower loss. This can intuitively be seen as the consequence of the emitter interacting longer with the photon before the latter escapes from the cavity, and gives rise to the well-known $Q/V$ dependence of the Purcell enhancement factor for the spontaneous emission rate [4], as first formulated by E. M. Purcell in 1946 for a system coupled to an electromagnetic resonator (here Q is the quality factor of the optical mode). In this case the spontaneous emission probability is increased over its bulk value, and the recombination time reduced, by a factor:

$$F_P = \frac{3\lambda_c^3}{4\pi^2}\frac{Q}{V} \qquad (2)$$

where $\lambda_c$ the wavelength in the material ($\lambda_c = \lambda_0/n_{ra}$, where $n_{ra}$ is the refractive index of the medium). Equation (2) applies only to the situation of a structure incorporating a spectrally narrow emitter, which in semiconductors implies negligible homogeneous broadening and therefore cryogenic temperatures. Interestingly, this enhancement applies also to stimulated emission – but again only in the ideal case of a narrow emitter [8]. In practical semiconductor lasers operating at room temperature, dephasing processes produce a homogeneous linewidth in the order of 10-20 meV (corresponding to 2.4-4.8 THz), typically much larger than the cavity linewidth (with the exception of high-loss plasmonic cavities). In this case, the integral in Eq. (1) simplifies to $L(\omega_{cav})$ and depends on the emitter's linewidth, not on the cavity linewidth.

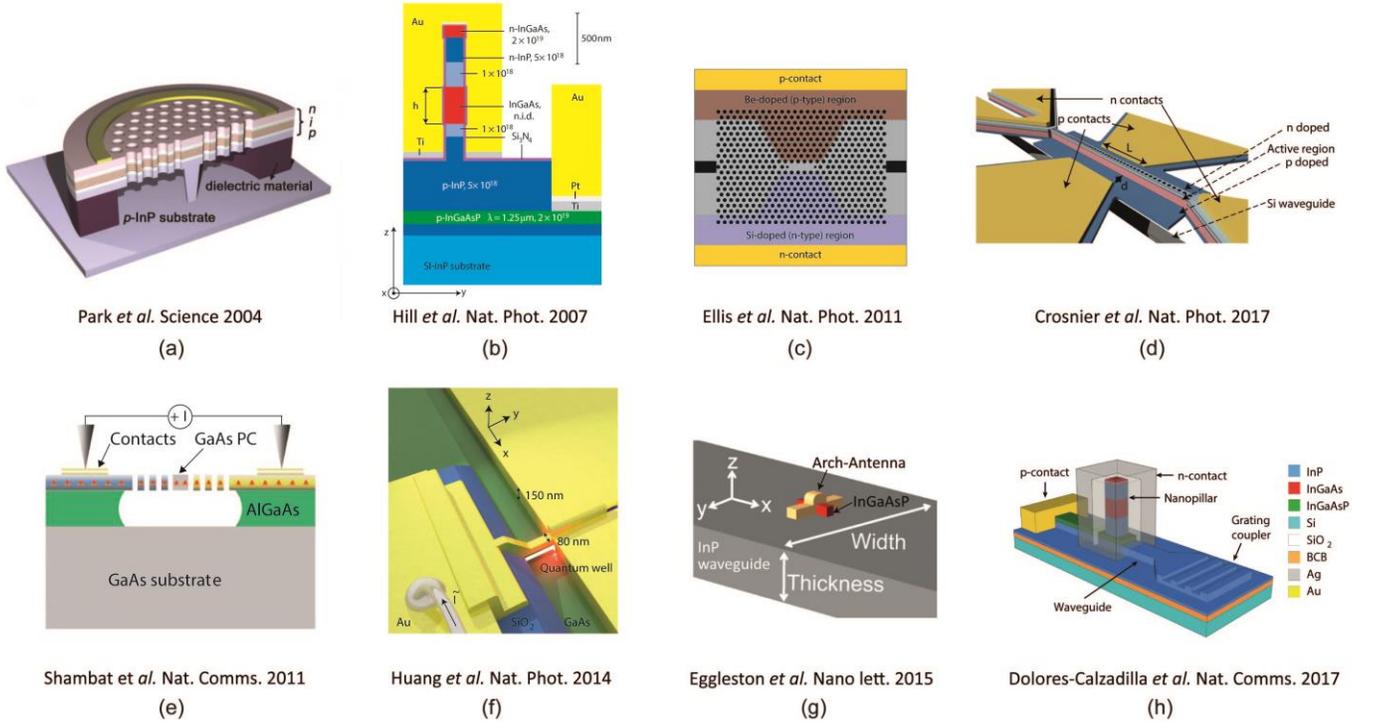

Fig. 2. A few representative structures of (a)-(d) nanolasers and (e)-(h) nanoLEDs using photonic crystal, metal-dielectric and plasmonic nanocavities. (a) Single-cell photonic crystal laser (reproduced from [10]), (b) Metallic-coated nanocavity laser (reproduced from [11]), (c) quantum-dot photonic crystal nanocavity laser (reproduced from [12]), (d) indium phosphide-on-silicon nanolaser (reproduced from [13]), (e) photonic crystal nanocavity light-emitting diode (reproduced from [14]), (f) nanoLED integrated with a deep-subwavelength plasmonic slot waveguide (reproduced from [15]), (g) optical antenna-enhanced nanoLED coupled to an integrated InP waveguide (reproduced from [16]), and (h) waveguide-coupled nanopillar metal-cavity LED on silicon (reproduced from [17]).

We note that these expressions in principle only apply to a localized quantum-confined gain material with negligible inhomogeneous broadening – for example an array of identical quantum dots all placed at a field maximum. For a more typical quantum well or bulk active region, an additional spectral integration over the bands and a spatial integration over the active region are needed [6], which tend to further reduce the rate enhancement as compared to the ideal case.

*B. Nanolaser and NanoLED Structures*

The impressive developments in the field of nanolasers in the last 10 years have been reviewed recently [3], [9], and will not be extensively described here. A few representative structures (but by far not exhaustive) are shown in Fig. 2 [10]–[17]. Restricting our attention to electrically-pumped devices, nanolasers can be classified in two broad categories, depending on the approach used to obtain tight optical confinement. One category consists of photonic crystal (PhC) lasers, where a wavelength-sized optical mode is defined through a defect in a photonic crystal (mostly a 2D PhC slab). PhC cavities can present a relatively high quality factor (~1000 or more [12]), enabling low-threshold operation, however efficient electrical injection and heat sinking are challenging. Also, the total footprint is significantly larger than the mode size, due to the presence of the PhC mirrors. In contrast, the optical confinement by metallic layers, employed in the second category, allows smaller (subwavelength) mode sizes, small total footprint, efficient electrical injection and heat sinking via the metal layer, but suffers from relatively high optical loss (quality factors typically in the few 100s [18]). So far, PhC lasers have shown the most promising performance, with threshold current in the µA range, differential efficiency >10%, operation at 10 Gb/s with energy budgets of few fJ/bit [19].

We note that metallic confinement does not necessarily imply a "plasmonic" character of the mode, which can quantified by the fraction of energy stored as kinetic energy of the free electrons [20], as plasmons polaritons are collective excitations of the free electron gas and the electromagnetic field. In many cases, the metal layer helps the confinement but the fraction of the energy in the metal is small and therefore the structure does not qualify as plasmonic. It was previously noted [20] that deeply-subwavelength plasmonic structures would present a threshold current independent of device size, leading to a diverging threshold current density in the limit of vanishing size. In practice, assuming maximum current densities in the order of 100 kA/cm$^2$ leads to a minimum lateral size of ~100 nm for a nanolaser (under the most optimistic assumptions for optical and carrier losses). In the following we will show that larger dimensions are in fact more likely to provide adequate performance for interconnects, due to the corresponding requirements on energy and speed.

In the field of nanoLEDs, the attention has been focused on scaling the mode volume aggressively in order to increase the spontaneous emission rate and thereby the efficiency and modulation speed to values exceeding those of nanolasers. The large spontaneous emission enhancement in metallic nanostructures, with internal quantum efficiencies exceeding 50%, i.e. $\tau_r^{-1}/(\tau_r^{-1} + \tau_{nr}^{-1}) > 0.5$, where $\tau_r$ is the radiative carrier lifetime, and $\tau_{nr}$ the non-radiative carrier lifetime, has been reviewed recently [21]. The investigated structures include nanoparticles or nanoantennas coupled to emitters (e.g. fluorescence dyes or quantum dots). Despite the remarkable results, including the impressive spontaneous emission lifetimes below 11 ps, anticipating ~90 GHz speeds [22], or an spontaneous emission rate enhancement of 115× in optical nanoantennas [23], in all these results only optically pumped structures are reported. Direct demonstrations of electrically modulated high-efficiency devices with speeds exceeding >10 GHz is still missing, due to the difficulty of combining ultrasmall mode volume and electrical injection.

A few representative structures of encouraging room-temperature high-speed LEDs are shown in Fig. 2(e)-(h). These include photonic crystal based LEDs showing 10 GHz modulation speed [14], although with output powers of the order of tens to hundreds of pW (at µW bias levels), i.e. an efficiency of ~10$^{-5}$, a waveguide-coupled nanoLED using a single-mode plasmon waveguide showing an efficiency ~10$^{-7}$ (here the modulation speed was not verified experimentally although a Purcell factor of 2 was estimated) [15]. Recently, a nanoLED based on a nanocavity photonic crystal cavity integrated with van der Waals heterostructures shows promising alternatives for planar nanoscale optoelectronics using 2D materials [24]. Despite the observed locally-enhanced electroluminescence, the modulation speeds are still low (MHz range).

Notwithstanding all these major advances, the reported output powers in these approaches have remained well below the nW level, partly due to the high losses and unoptimized waveguide coupling. In 2017, a waveguide-coupled nanopillar metal-cavity LED was reported by us and colleagues [17]. The nanoLED consisted of a semiconductor nanopillar (lateral size ~300 nm) encapsulated with metal, evanescently coupled to a low-loss InP waveguide on a silicon substrate. An on-chip external quantum efficiency between 0.01% and 1% was obtained at 300 K and 9.5 K, respectively, corresponding to waveguide-coupled power levels of around 20 nW and 300 nW. The room temperature (RT) efficiency was strongly limited by nonradiative recombination at the nanopillar surface. This fast nonradiative recombination was beneficial to the device speed, and indeed electrical-optical modulation experiments revealed that the nanoLED converts electrical signals into optical signals at rates up to 5 Gb/s. In more recent work, our group has reported ultralow surface recombination velocities of ~260 cm/s in InGaAs/InP undoped nanopillars [25], which, as discussed below, would significantly improve the LED performance.

Although all these recent results create wide expectations for high-density nanoLED/nanolaser-based optical communication systems at Gbps data rates using ultra-low power consumption, in the following we examine some crucial energy/speed limits related to the downscaling of these sources, which are key for the future design of practical nanoLEDs and nanolasers.

## III. Size scaling of nanoLEDs

NanoLED sources are in principle ideal for applications in very-short-distance on-chip or chip-to-chip communications. Their potential advantages, as compared with nanolasers, include:

*(i) compatible with high-loss cavities;*
*(ii) operate without a threshold, hence higher efficiency at low injection;*
*(iii) potentially less complex fabrication and higher yield;*
*(iv) less complex driving circuitry and potentially higher thermal stability due to the absence of threshold.*

Notably, while the efficiency of conventional LEDs is limited by the fact that only a fraction of spontaneously emitted photons can be collected, the modification of the spontaneous emission rate occurring in a wavelength-sized cavity can in principle provide an avenue to the control of the emission process. Indeed, as discussed in Section II.A, the spontaneous emission rate scales as the inverse of the mode volume, and, in the case of a spectrally narrow emitter, with the quality factor of the cavity, leading to large enhancements for small and low-loss cavities. This has led to the expectation that nanoLEDs could be both efficient and fast. However, as discussed in Section II, the homogeneous broadening, inevitable in a semiconductor active medium at room temperature, significantly reduces this enhancement. Additionally, the distribution of carriers in the band must be considered. As a consequence, achieving radiative enhancements in electrically modulated nanoLEDs and respective speeds comparable or larger than laser devices (e.g. > 20 GHz) can be challenging. Even in the case of plasmonic cavities with broad emitters where these potential fast speeds can be achieved as a result of large Purcell factors due to extremely small mode volumes [21], achieving these speeds with sufficient output power levels still needs to be realized. As shown below, this is in fact unlikely due to practical limitations on the current density.

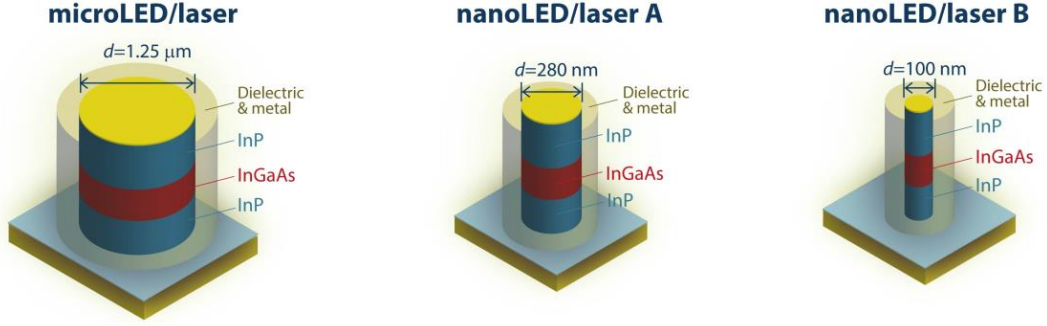

Fig. 3. Schematic of representative micro- and nanoscale LEDs and lasers. The structures are examples of metallo-dielectric cavities which can confine light to volumes with dimensions smaller than the wavelength. They typically consist of a semiconductor pillar using a double heterostructure InP/InGaAs/InP, surrounded by an isolating dielectric material and then encapsulated with metal. The combination of metal and dielectric confines the optical mode around the semiconductor gain region.

Here, we describe the single-mode rate-equation model considering the realistic situation of a nanoLED operating at room-temperature and employing a bulk active medium with homogeneous broadening larger than the cavity linewidth and inhomogeneous broadening of the electronic states. The inhomogeneous broadening must be described by integrating Eq. (1) over the bands. In the limit $\Delta\omega_{em} >> \omega_{cav}$, Fig. 1, the spontaneous emission rate per unit time and volume, $r_{sp,cav} = R_{sp,cav}/V_a$ (where $V_a$ is the volume of the active material), becomes [6]:

$$r_{sp,cav} = \frac{\pi}{\hbar\varepsilon_0\varepsilon_{ra}} d_{if}^2 \frac{\omega_{cav}}{V}$$

$$\times \int_{\omega_g}^{\infty} \rho_j(\omega_{vc}) L(\omega_{cav} - \omega_{vc}) f_c (1 - f_v) d\omega_{vc} \quad (3)$$

where $\rho_j(\omega_{vc})$ is the joint density of states per unit frequency and volume, and $f_c$, $f_v$, are the Fermi distribution functions of electrons in the conduction band and valence bands, respectively. As $r_{sp,cav}$ is inversely proportional to the mode volume, we can also define a volume-independent parameter, $\gamma_{sp,cav}$, $r_{sp,cav} \equiv \frac{\gamma_{sp,cav}}{V}$ to explicitly show the volume dependence in the rate.

Using (3) describing the photon creation rate by spontaneous emission in a resonant cavity we can write the rate equations for carrier density, $n$, and photon density, $n_{ph}$, to describe an electrically modulated nanocavity LED:

$$\frac{dn}{dt} = \frac{\eta_i I}{qV_a} - r_{nr} - r_l - \frac{\gamma_{sp,cav}}{V_{eff}} \quad (4)$$

$$\frac{dn_{ph}}{dt} = \frac{V_a}{V_{eff}} \frac{\gamma_{sp,cav}}{V_{eff}} - \frac{n_{ph}}{\tau_p} \quad (5)$$

where $I$ is the injection current, $q$ is the electron charge, and $\eta_i$ the injection efficiency, $r_{nr} = \frac{\upsilon_s A}{V_a} n + Cn^3$ describing the rate of nonradiative recombination, that accounts for the surface recombination (described by the surface velocity, $\upsilon_s$, and by the surface area of the active region, $A$), and for Auger recombination, $C$, and $r_l$ describes the radiative decay into all other modes (or leaky modes) [26]. The remaining parameters include $\gamma_{sp,cav}/V_{eff}$, the spontaneous recombination rate term, where $V_{eff}$ is the effective mode volume, which replaces $V$ in (3) to take into account a spatially distributed emitter, see more details in [6], $V_a/V_{eff}$ that can be defined as the confinement factor [7], $\Gamma$, and the term $n_{ph}/\tau_p$ which denotes the photon escape rate determined from the cavity $Q$-factor (where $\tau_p = \lambda_0 Q/2\pi c$ is the photon lifetime). We note in (5) that the spontaneous emission term has a $1/V_{eff}^2$ dependence, where the additional $1/V_{eff}$ dependence is a result of the use of volume densities in the rate equations. In (4)-(5) the rate of stimulated emission is assumed to be negligible and it is not included in the model. For purposes of numerical simulation, the spontaneous emission rates are calculated as a function of Fermi levels and the carrier population is retrieved from the charge neutrality condition, see details in [6].

The rate-equation model presented here is quite general and can be used to analyze both micro- and nanoscale LED sources. It is important to note that this model avoids the *ad-hoc* introduction of the Purcell factor directly into the rate equations. This is important as only the physical parameters of the nanoLED, such as the cavity dimensions and emitter/cavity relative linewidths, can be designed in a realistic nanocavity LED. Lastly, we note that our goal is not to provide an extensive model that includes all relevant effects (e.g. temperature effects, band nonparabolicity), but rather an intuitive physical description of practical nanoLED structures under realistic operating conditions.

In what follows, departing from the simple rate-equation model, (4)-(5), and using parameter values relevant for recently reported nanoLEDs [17], [25] and nanolasers [11], [27], we examine the various scenarios of the role of radiative and non-radiative recombination, specifically surface recombination and radiative enhancement, in nanoLEDs in terms of energy/speed limits.

*A. NanoLED light-current characteristics*

In this subsection, our aim is to analyze the effect of radiative and nonradiative recombination, and specifically surface and Auger effects, in the efficiency of nanoLED sources using realistic parameters and practical operation at room-temperature. For this purpose, using the rate-equation in (4)-(5), we simulate the light-current ($L-I$) characteristics of three representative LED nanostructures, a microLED, a nanoLED A and a nanoLED B, shown schematically in Fig. 3. In order to make the discussion concrete, we consider structures having similar designs as the ones presented in various works (e.g. [11], [17]) and consisting of metallo-dielectric cavities made of a pillar-like semiconductor active region (e.g. InP/InGaAs/InP double heterostructure) surrounded by an isolating dielectric material and then coated with metal. In this work we focus our study on a bulk InGaAs active material. For a complete description of the typical parameters used in the model for the InGaAs bulk active material see our recent work [6]. We note, the main conclusions reported here for the bulk case should still be valid for a case of multiple quantum wells (MQWs) where also the inhomogeneous and homogeneous broadenings typically overwhelm the cavity linewidth. The disadvantage of using MQWs, namely in nanolasers, is related with the lower modal gain that can be achieved as compared with the bulk material [28]. The same structures have been analyzed by the authors in a recent work to study the static and dynamic properties of micro- and nanolasers [6]. While we focus on specific structures, we note that the approach is completely general and applies to a wide range of other nanophotonic cavities, including for example photonic crystals.

For practical analysis and direct comparison, we assume identical quality factor of $Q=60$ for all cavities (corresponding to a photon lifetime of $\tau_p = 49$ fs), that is, much smaller than the typical quality factors >200 needed to achieve lasing [27]. For simplicity of analysis, we assumed an effective volume scaling with the physical volume of the InGaAs active disk (height of 300 nm) and keep $V_a \sim 0.8 V_{eff}$. Therefore, the effective mode volume varies from 0.5 μm$^3$, microLED case, to 0.025 μm$^3$, nanoLED A, and lastly to 0.0025 μm$^3$, nanoLED B. We note that, even in the smallest nanoLED B, the cavity dimension is still larger than the wavelength in at least one direction. Therefore, more energy is stored in the magnetic field than in the motion of electrons in the metal, as described in [29], so that the assumptions used here for the definition of effective mode volume are still valid. In a plasmonic case, the kinetic energy needs to be included in the calculation of the effective mode volume, as discussed in [29]. Lastly, the calculated photon density in (4)-(5) was converted to an output power using:

$$P = \frac{n_{ph} V_{eff} \eta hc}{\tau_p \lambda_0} \tag{6}$$

where $h$ is the Planck's constant, and $\eta$ the coupling efficiency (useful loss/total loss). For simplicity of analysis, both injection and coupling efficiencies were kept constant, and set to $\eta_i = \eta = 1$. We note that approaching this ideal limit will require an ideal diode where leakage current effects can be neglected and careful design of the coupling between cavity and output waveguide – promising progress in this direction has been reported [17], [19], [30].

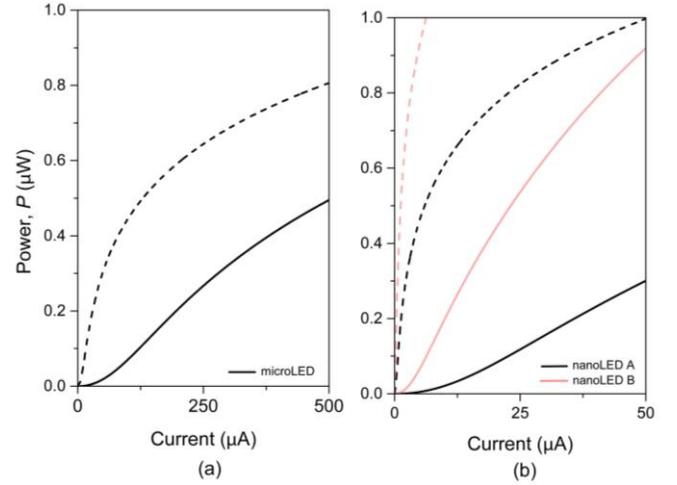

Fig. 4. Simulated $L-I$ characteristics of the (a) microLED, and (b) nanoLED A and nanoLED B. In all curves the solid lines correspond to a value of surface recombination velocity of $7\times10^4$ cm s$^{-1}$, and the dashed lines to 260 cm s$^{-1}$.

In the calculations we will particularly consider the range of output powers of 1 to 10 μW, as these correspond to an optical energy per bit of 1 fJ at 1 Gb/s and 10 Gb/s, respectively. Indeed, as mentioned above, data communication with thermal noise-limited receivers requires at least 1 fJ/bit considering some margin for loss. Figure 4 displays the calculated $L-I$ curves showing the optical power versus the injected current for (a) the microLED and (b) nanoLEDs A and B. The curves were simulated considering the following values for the surface recombination: a large surface velocity value of $7\times10^4$ cm s$^{-1}$ (continuous lines), typically found in micro- and nanoLED/laser devices [17], [27], and an ultralow value of surface recombination of 260 cm s$^{-1}$ (dashed lines), achieved recently in InGaAs/InP nanopillars using an improved passivation method [25]. In all plots, we kept a realistic room-temperature Auger coefficient [6]. Firstly, in the case of the microLED with a mode volume of 0.5 μm$^3$ and assuming a large surface recombination, solid line in Fig. 4(a), a current injection as high as 500 μA is needed to reach output power levels around 0.5 μW. The level of current injection for the same output power reduces by almost five times when assuming a strong reduction of the surface recombination. However, achieving power levels >1 μW at low current injections, and thereby high efficiency, remains challenging using these microLED structures. Indeed, the spontaneous emission rate in the mode of interest is low due to the relatively large mode volume, and most of the emission couples to leaky modes (we use a rate of $r_l = 2\times10^{-8}$ s$^{-1}$, a value lower than the emission in a bulk material which assumes the suppression of the emission of the leaky modes typical of nanophotonic cavities (see, e.g. in micropillar cavities [31]). We note that the single-mode efficiency calculated here applies to LEDs coupled to a single-mode output channel (e.g. an on-chip waveguide), as appropriate for high-data rate communication, whereas the efficiency for free-space coupling can be much higher.

We now analyze the cases of nanoLED A and B with mode volumes of 0.025 μm$^3$ and 0.0025 μm$^3$, respectively, Fig. 4(b). When assuming a large surface recombination velocity, power levels close to 1 μW at only 50 μA of injected current can already be achieved for the case of nanoLED B, and also for the case of the nanoLED A for current levels >> 50 μA (not shown in the plot). The increased efficiency for lower mode volumes is directly related to the increase of the spontaneous emission rate in the mode, $R_{cav} \propto 1/V$. However, the enhancement of the total spontaneous emission rate remains limited: The Purcell factors calculated for nanoLED A and B are in the order of 1 and 10, respectively [6]. This is due to the large homogeneous and inhomogeneous broadening, as explained in Section II. We note that several previous works that analyze InGaAsP bulk [32], quantum well [15], [26], emitters already recognize that achieving large Purcell factors (>10) in sub-wavelength cavity lasers/LEDs is challenging when pronounced broadening effects are taken into account. We note that it is in principle possible to achieve high Purcell factor also in the presence of large broadening, but only for structures <<λ where the current density limitation affects the total output power and bandwidth, as discussed in Section IV. Since we assume $\eta_i = \eta = 1$, and thermal effects are not considered in our analysis, the calculated power levels clearly correspond to the best-case scenario. Therefore, experimentally achieving power levels above 1 μW in practical nanoLEDs exhibiting strong surface nonradiative rates, is extremely challenging. On the other hand, when a reduction of the surface recombination to a value of 260 cm s$^{-1}$ is considered [dashed lines in Fig. 4(b)], our simulations suggest a substantial improvement of the predicted efficiency of the nanoLEDs, corresponding to a 100-fold increase of the output power at low current injections. For example, in the case of the Purcell-enhanced nanoLED B, a current of only ~10 μA is needed to reach an optical output of 1 μW. We note that for the extremely low surface velocities considered here, the main source of inefficiency is emission into leaky modes. Notwithstanding the impressive improvement, we note the output power at higher injection current levels begins to saturate due to the unavoidable Auger recombination effect. This does not include further saturation effects due to, e.g. temperature, which are known to strongly affect the performance of nanoLEDs at room temperature [17]. Indeed, a current of 10 μA corresponds to a current density of 16 kA/cm$^2$ in nanoLED A and 127 kA/cm$^2$ in nanoLED B, which in the presence of any series resistance will lead to strong local power dissipation and heating. As an example, for the smallest pillar even the resistance of a 100-nm thick p-doped InP already provides a large contribution of ~2.5 kΩ to the total device resistance (assuming a resistivity of 2×10$^{-2}$ Ω.cm and doping of 10$^{19}$ cm$^{-3}$). Together with further contributions from the n-doped InP and metal contacts this will likely lead to unsustainable heat generation making operation >100 kA/cm$^2$ extremely challenging.

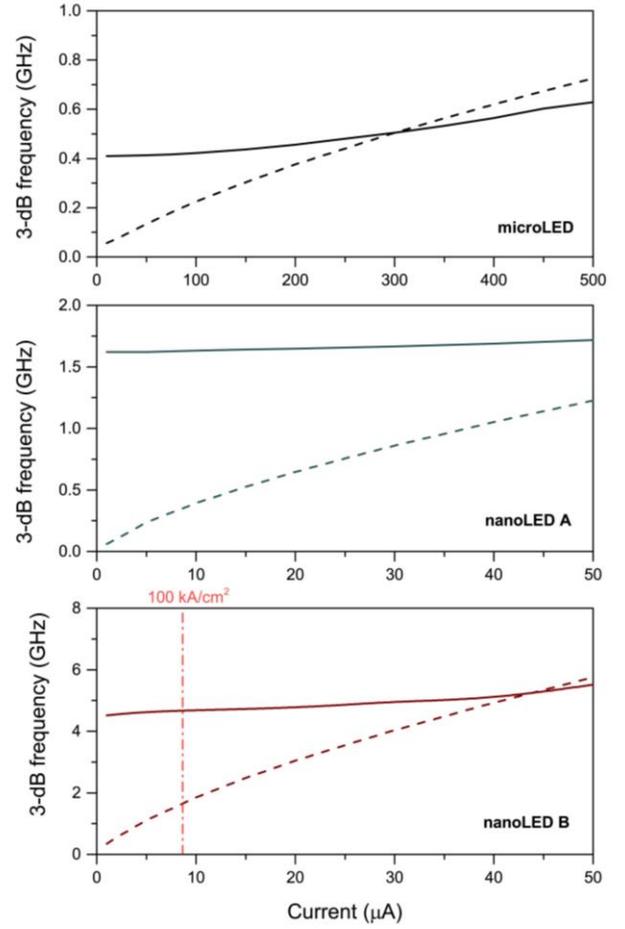

Fig. 5. Simulated small-signal 3dB bandwidth versus current injection for the microLED, nanoLED A and nanoLED B devices analyzed in Fig. 4. In all curves the solid lines correspond to a value of surface recombination velocity of 7×10$^4$ cm s$^{-1}$, and the dashed lines to 260 cm s$^{-1}$. The dashed-dot vertical line in nanoLED B plot indicates the current corresponding to a current density of 100 kA/cm$^2$ (in the remaining cases the micro/nanoLEDs operate below this limit).

In summary, we have identified some of the crucial key parameters required to examine the performance of nanoLEDs. Our simulations scenarios clearly suggest that highly-efficient nanoLEDs at room-temperature with power levels >1 µW and current injections <<100 µA, are in principle possible. This however corresponds to the best-case scenario of Purcell enhanced LEDs when optical and carrier losses including nonradiative effects can be strongly mitigated. Practical considerations on maximum current density are likely to limit the output power to well below the 1 µW level.

*B. NanoLED modulation bandwidth*

Here we analyze the effects of the mode-volume scaling, surface recombination and Auger recombination on the modulation speed. To obtain the high-speed modulation response, we perform a small-signal analysis of equations (4)-(5) following a standard procedure described in detail in [6], and calculate the 3dB modulation frequency as a function of injection current for the three structures (Fig. 5).

The simulations results of both $L-I$ curves and signal 3 dB bandwidth versus current injection, Figs. 4 and 5 respectively, show a clear compromise between speed and efficiency for all analyzed LED sizes. This is observed mainly at low injection currents and is related to the effect of surface recombination. For large surface velocity values (solid lines in Fig. 5), a high-speed modulation bandwidth (> 1 GHz) can be achieved for both nanoLEDs A and B with current injections of only 10 µA. Nevertheless, the corresponding output power levels are <100 nW which is too low for data communication. A further increase of the injection current does not significantly change the modulation speed (in the current ranges analyzed here). However, the 3-dB bandwidth curves dramatically change when assuming a low surface velocity. As shown in the dashed curves of Fig. 5, when a surface velocity of 260 cm s$^{-1}$ is considered, the modulation speed at low current becomes strongly dependent on the radiative recombination rate. The modulation bandwidth increases for smaller devices (Fig. 5 middle and bottom panel). This is due to the increased effect of surface recombination, particularly for the case of high surface recombination velocity (continuous lines), and also to the increase radiative rate for the smallest nanoLED B. In both cases the bandwidth is independent of the injection level for high surface recombination velocity (due to the monomolecular nature of surface recombination), while it increases with injection in low surface recombination case, due to radiative and Auger recombination. This in principle enables also the devices with low surface velocity values to reach bandwidths 1 GHz (nanoLED A) and 5 GHz (nanoLED B) at currents of 50 µA. However, these values correspond to extremely high current densities, particularly for the smallest device. Limiting the current density to 100 kA/cm$^2$ (~60 µA for nanoLED A and ~8 µA for nanoLED B) results in a maximum bandwidth of <2 GHz for both devices.

In summary, Purcell-enhanced nanoLED devices in the best-case scenario, specifically small devices with the geometry of nanoLED B and low surface passivation, can potentially operate at room temperature at 1-2 Gb/s data rates with on-chip optical power levels slightly above 1 µW (corresponding to

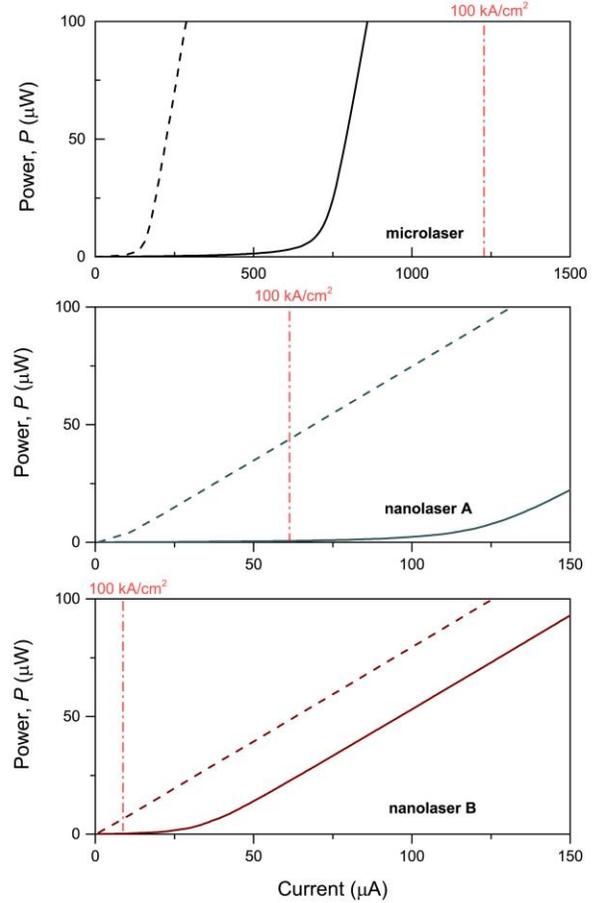

Fig. 6. Simulated *L-I* characteristics of the microlaser, nanolaser A and nanolaser B. In all curves the solid lines correspond to a value of surface recombination 7×10$^4$ cm s$^{-1}$, and the dashed lines to 260 cm s$^{-1}$. Also shown for each case in a vertical dashed-dot red line is the corresponding current density value of 100 kA/cm$^2$.

~10$^4$ photons/bit at 1 Gb/s) and with energy consumptions <10 fJ/bit (operation at 10 µA and assuming a voltage drop of 1.0 V), in the range required by on-chip optical interconnects. It should be emphasized that reaching significantly higher bandwidths without strongly compromising the efficiency will be exceedingly difficult in a practical semiconductor LED, due to the limited available Purcell effect related to linewidth broadening. A further downscaling below the 100 nm limit, while increasing the Purcell enhancement, would result in limited emission power. For example, even in highly unlikely case of perfect electro-optical conversion, a device with lateral dimension <100 nm would need to operate at current densities >100 kA/cm$^2$ to produce the 10 µW output power required for 10 Gb/s operation. Therefore, additional methods that do not require Purcell enhancement to further improve the speed of the nano-LEDs may be crucial in future designs to enable high-bandwidth operation of nanoLEDs. This includes taking advantage of Auger recombination at high carrier densities, as shown in our simulations, or using reverse-biasing of the nano-LED during the turn-off cycle to shorten the minority carrier storage time, as experimentally demonstrated in [17].

## IV. Size scaling of nanolasers

### A. Purcell effect and enhanced stimulated emission

Here, we describe the characteristics of the microlasers and nanolasers introduced in the previous section, which can be described by a set of single-mode rate equations rigorously derived in [6]:

$$\frac{dn}{dt} = \frac{\eta_i I}{qV_a} - r_{nr} - r_l - \frac{\gamma_{sp,cav}}{V_{eff}} - \gamma_{net} n_{ph} \qquad (8)$$

$$\frac{dn_{ph}}{dt} = \frac{V_a}{V_{eff}} \gamma_{net} n_{ph} + \frac{V_a}{V_{eff}} \frac{\gamma_{sp,cav}}{V_{eff}} - \frac{n_{ph}}{\tau_p} \qquad (9)$$

These equations are similar to (4)-(5) used to analyze the nanoLEDs. In this case however we include the net stimulated emission per unit time and volume,

$$\gamma_{net} n_{ph} = \frac{\pi}{\hbar \varepsilon_0 \varepsilon_{ra}} d_{if}^2 \omega_{cav} n_{ph}$$

$$\times \int_{\omega_g}^{\infty} \rho_j(\omega_{vc}) L(\omega_{cav} - \omega_{vc})(f_c - f_v) d\omega_{vc} \qquad (10)$$

This expression, apart from the different role of the occupation probabilities and for the presence of the photon number, is identical to the one obtained for the spontaneous emission into the cavity mode, (3). Clearly, for a given mode, both stimulated and spontaneous rates are enhanced exactly by the same factor, as it is expected from Einstein's relations. We stress again that the $1/V_{eff}$ factor in the gain term, which also appears in the standard laser equations [7], comes from the increased field per photon in smaller cavities. It should be noted that in (10) we have not considered the reduction in differential gain [7] and gain compression [33] (that is, $\gamma_{net}$ does not depend on $N_{ph}$), which may have a role in the dynamics of nanolasers, specifically in the modulation response.

### B. Nanolaser static and dynamic properties: peculiar effects

In Fig. 6 the simulated $L-I$ curves using the rate-equation model (8)-(9) are displayed, showing the optical power versus the injected current considering both cases of high and low surface recombination velocities. The curves were calculated using the same parameter values as used previously for the nanoLED cases, except for the quality factor. Here, we assumed a quality factor $Q = 235$ for all cavities in order to achieve lasing in practical structures, corresponding to a photon lifetime of $\tau_p = 0.19$ ps. In practical structures, $Q > 200$ can be achieved at room temperature using optimized metal layers [34]. In all plots, the current value corresponding to a current density of 100 kA/cm² is shown as a vertical red dashed-dot line as a reference where temperature effects become relevant in metallic cavities [17]. Considering first the case of the microlaser in Fig. 6, lasing can be achieved at <100 kA/cm² using both large and low surface recombination velocity values. For this choice of parameters, lasing is achieved at a threshold current density of ~56 kA/cm² for 7×10⁴ cm s⁻¹ and is 5-fold reduced to ~10 kA/cm² considering a strong reduction of the surface velocity to 260 cm s⁻¹. For both cases, an optical power level well above 10 μW can be reached at reasonable current densities. As expected, lasers enable producing much larger powers than LEDs, as a consequence of the increased stimulated emission rate at higher injections, while the population remains (partly) clamped. This is fundamental advantage over LEDs, where increasing the injection inevitably leads to increased Auger recombination, even when the low-injection efficiency is high.

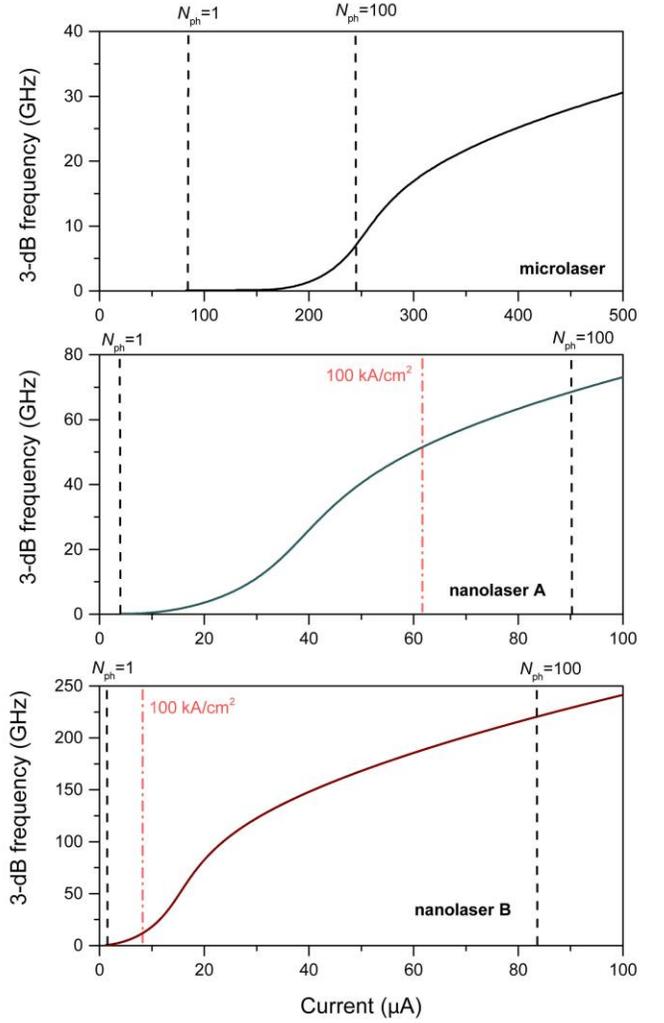

Fig. 7. Simulated small-signal 3dB bandwidth versus current injection for the microlaser, nanolaser A and nanolaser B devices analyzed in Fig. 6 considering the best case scenario of low surface recombination velocity value of 260 cm s⁻¹. The vertical dashed lines show the corresponding photon number and the vertical dashed-dot red lines indicate the current density limit value of 100 kA/cm².

Analyzing the $L-I$ curves of nanolasers A and B, when assuming a large value surface velocity, neither of the nanolasers achieve threshold below the 100 kA/cm² limit considered here. However, in the case of low surface velocity, lasing can be achieved at threshold current densities of ~7 kA/cm² and ~15 kA/cm² for nanolasers A and B, respectively. This calculation clearly demonstrates the key role of surface recombination velocity in the performance of sub-μm laser structures operating a room-temperature. Note that, if we keep the current density below 100 kA/cm², nanolaser B reaches an

10output power level of only <10 µW. Clearly, operating ultrasmall lasers at tens of µW with sustainable current densities remains a challenge. Another key effect that can be observed in Fig. 6 for nanolasers A and B when the nonradiative losses are low is the smooth transition from non-lasing to lasing, resulting in a "thresholdless" behavior, particularly for the case of nanolaser B. For the parameters used here, this effect is a direct result of the substantial reduction of the surface recombination rate together with the small mode volume which results in a substantial fraction of the spontaneous emission coupled to the cavity mode below threshold. The calculated spontaneous emission factor, $\beta = R_{sp,cav}/(R_{sp,cav} + R_l)$, indeed indicates values close to unity for the case of nanolaser B and explains the "thresholdless" behaviour in this case. The β factor (not shown) varies from around 0.01 for the case of the micropillar laser up to ~0.8 in the case of the smallest nanopillar laser B (assuming in both cases a carrier density of $10^{18}$ cm$^{-3}$). Here and in the following, we define as threshold the bias point for which $N_{ph} = 1$, i.e. the point from which stimulated emission starts to dominate.

Stimulated emission obviously also makes nanolasers faster. In Fig. 7 we show the calculated small-signal 3dB-bandwidth as a function of the injected current plotted for $N_{ph} > 1$ considering the best-case scenario of a nanolaser with low surface recombination velocity. The dashed lines show the photon number corresponding to a given current and device size, and the vertical dashed-dot lines indicate the current density limit of 100 kA/cm$^2$. Analyzing the speed at a constant photon number, $N_{ph} = 100$, the 3dB-bandwidth clearly shows a large increase from well below 10 GHz for the microlaser to >50 GHz for nanolaser A, and >200 GHz for nanolaser B. This allows us to conclude that a large increase of speed in nanolasers can be achieved as a direct consequence of the strong reduction of the effective mode volume and corresponding enhancement of the stimulated emission rate.

An insight in the modulation characteristics of nanolasers can be obtained from the expressions of the relaxation oscillation frequency and damping rate. As rigorously derived in [6], in the situation where i) the nonradiative contribution can be neglected, ii) $V_{eff} \sim V_a$, as the cases considered here, and iii) the contribution of the leaky modes can be neglected (e.g. in the case where the radiative emission into the the lasing mode is large), the relaxation frequency, $\omega_R$, and the damping factor, $\gamma_R$, can be approximated as:

$$\omega_R^2 \approx \frac{N_{ph}}{\tau_p V_{eff}} \frac{\partial \gamma_{net}}{\partial n} + \frac{1}{\tau_p V_{eff}} \frac{\partial \gamma_{sp,cav}}{\partial n} \quad (11)$$

$$\gamma_R \approx \tau_p \omega_R^2 + \frac{1}{\tau_p} - \gamma_{net}(n_0) \quad (12)$$

where $n_0$ in $\gamma_{net}(n_0)$ in (12) is the stead-state carrier density value. For the cases analyzed here, when $N_{ph} > 10$ the relaxation oscillation frequency can be further simplified to $\omega_R^2 \approx \frac{N_{ph}}{\tau_p V_{eff}} \frac{\partial \gamma_{net}}{\partial n}$, which agrees with the typical expression found in laser textbooks [7]. The dependence of $\omega_R$ on the inverse of $V_{eff}$ shows how the modulation dynamics is not affected by the spontaneous emission term and depends only on the module volume though the gain term, as expected from a typical laser source. Only in the case of very low-photon regime (typically $N_{ph} < 10$ for the cases analyzed here) the spontaneous emission plays a role and the full expression in (11) needs to be considered to entirely describe the relaxation oscillation frequency.

In the expression (12) of the damping rate, also two regimes can be distinguished. For low photon numbers (ranging from $N_{ph} = 1$ to $N_{ph} = 20$), the factor $\tau_p \omega_R^2$ in (12) can be neglected and the damping factor depends approximately on $\frac{1}{\tau_p} - \gamma_{net}(n_0)$. In this case, a large decrease of the damping with injection (more than one order of magnitude) can be observed, due to the increasing gain. This makes the current dependence of the damping factor in nanolasers markedly different form the one in large lasers where the $\frac{1}{\tau_p} - \gamma_{net}(n_0)$ term can be assumed to be zero and the damping increases with the current injection following $\tau_p \omega_R^2$. This increase in the damping rate can also be observed in nanolasers, but only at photon numbers corresponding to extremely large current densities (>>200 kA/cm$^2$). Realistically, deep-subwavelength lasers with low-Q factors, as the ones analysed here, will always operate in an overdamped regime. The effect of overdamped relaxation oscillations in nanolasers based on few discrete emitters has been also described in the work of Moelbjerg *et al.* [35].

Apart from these peculiar effects, namely in the damping factor, we note that the advantage of deep-subwavelength scale nanolasers (≤100 nm scale) in terms of modulation speed is not entirely obvious. For example, keeping the current density limit of 100 kA/cm$^2$ as a realistic condition, the maximum bandwidth of nanolaser B is strongly compromised, as it does not operate much above the threshold. In fact, considering this limit, nanolaser A could have modulation bandwidths close to 50 GHz with power levels >25 µW while nanolaser B would be limited to slightly above 10 GHz with an optical power <10 µW. Only at extremely high, and probably unsustainable, current densities of >250 kA/cm$^2$, one could obtain a modulation speed >100 GHz for nanolaser A. As a result, a clear trade-off between speed, efficiency and current density is reached when the cavity size of nanolasers is reduced to deep sub-wavelength regimes, i.e. ≤100 nm scale. Lastly, we note that nanolaser B biased below 250 kA/cm$^2$ works in a region unusual for standard lasers, where both spontaneous and stimulated emission into the cavity mode significantly contribute to the emission. In this region, the phase and intensity noise properties of the laser can be affected [36], [37], which can impact the performance of the laser in a communication system.

Furthermore, we note the analysis considered here is an optimistic best-case scenario. We assumed the nanolasers were operating under ideal, and therefore unlikely, conditions of maximum injection and coupling efficiencies, negligible heating, and no gain compression effects.

*C. Scaling scenarios: nanoLEDs vs. nanolasers*

In what follows, using the simulations of nanoLED and nanolaser sources presented in the previous sections, we summarize the potential performance of these sources in terms of energy efficiency and ask ourselves which type of source is most suited to ultralow-power optical communications.

In Fig. 8 we plot the needed electrical energy as a function of the optical energy per bit, for both micro- and nanoLED/lasers operating at data rates of 1 and 10 Gb/s. For all plots, we considered the best case scenario of a nanoLED/laser with a low surface recombination velocity. The curves were plotted in the regions where the modulation bandwidths are larger than the corresponding bit rate (assuming the available modulation bit rate is 1.3 times higher than the 3-dB small signal bandwidth [38]). For the case of the nanoLEDs, only the plots at 1 Gb/s are shown [Fig 8(top)] since, as discussed in Fig. 5, the LEDs analyzed here can only operate <<10 GHz. The optical energy is calculated as $P \cdot T$, where $T$ is the bit duration while for the electrical energy we assume $I \cdot V \cdot T$, optimistically assuming $V$=1.0 V (i.e. neglecting the increase in series resistance to be expected in smaller structures). In Fig. 8, the curves are only plotted in the regions where LED/laser devices operate with a current density <100 kA/cm$^2$. We note that the ratio optical to electrical energy is also equal to the laser/LED wall-plug efficiency (WPE), which is defined by the ratio of emitted optical power to consumed electrical power.

For the cases of the micro and nanoLEDs (dashed lines) shown in Fig. 8(top), the only potential LED structure of interest for energy-efficient optical data communications is nanoLED B. This deep-subwavelength device can operate at 1 Gb/s using energies below 10 fJ/bit while providing optical energies slightly above 1 fJ. However, the nanoLEDs analyzed here are limited at data rates exceeding only a few Gb/s and the expected optical power levels > 1 μW correspond to a best case scenario. Furthermore, they require to operate ~100 kA/cm$^2$. NanoLEDs clearly reach a fundamental energy/speed limit for data rates exceeding 1 Gb/s.

For the case of the micro- and nanolasers, the expected performances reveal interesting potential for low-energy short-distance optical communications at high speeds. In the case of the microlaser, this device size is well suited for operation in the range of 10-100 fJ/bit at 10 Gb/s and with optical energies well above 10 fJ. This device structure indeed resembles a scaled-down version of a vertical-cavity surface-emitting laser (VCSEL), with decreased area and increased thickness of the active region to compensate for the higher optical loss. Its performance is comparable to the one of the smallest oxide-confined VCSELs (13.5 Gb/s with 0.97 pJ/bit energy efficiency [39])

In the case of the sub-μm nanolasers A and B, the simulations reveal an improved performance at low energy. Nanolaser A, due to its threshold-less behavior, can operate – in this idealized scenario - with nearly unity energy (and wall-plug) efficiency at energy levels >1 fJ and data rates of 10 Gb/s. This nanolaser structure can also operate at modulation rates up to 40 Gb/s (not shown) at <10 fJ/bit energy levels while maintaining sustainable current density levels <100 kA/cm$^2$. However, its competitive advantage with respect to the larger microlaser becomes less evident at energies >>10 fJ/bit, as at larger power levels also the threshold power of the latter becomes negligible. In the case of the deep-subwavelength size nanolaser B, only the operation at 1 Gb/s is promising for this small size laser, shown in Fig. 8(top). Operation at data rates of 10 Gb/s requires current densities >>100 kA/cm$^2$ in order to produce optical energies above 1 fJ/bit, and corresponding strong heating effects which will likely compromise its performance. These results also agree with the main conclusions of the work of Khurgin *et al.* [29] for the case of subwavelength plasmonic lasers (spasers), reporting modulation speeds of hundreds of gigahertz, but only at extremely high current densities of 10 MA/cm$^2$. In summary, for the cases analyzed here nanolasers with active dimensions in the range of few 100s nm (i.e. similar to nanolaser A) may enable the best and most flexible performance for high-density, short-distance and high-speed optical communications.

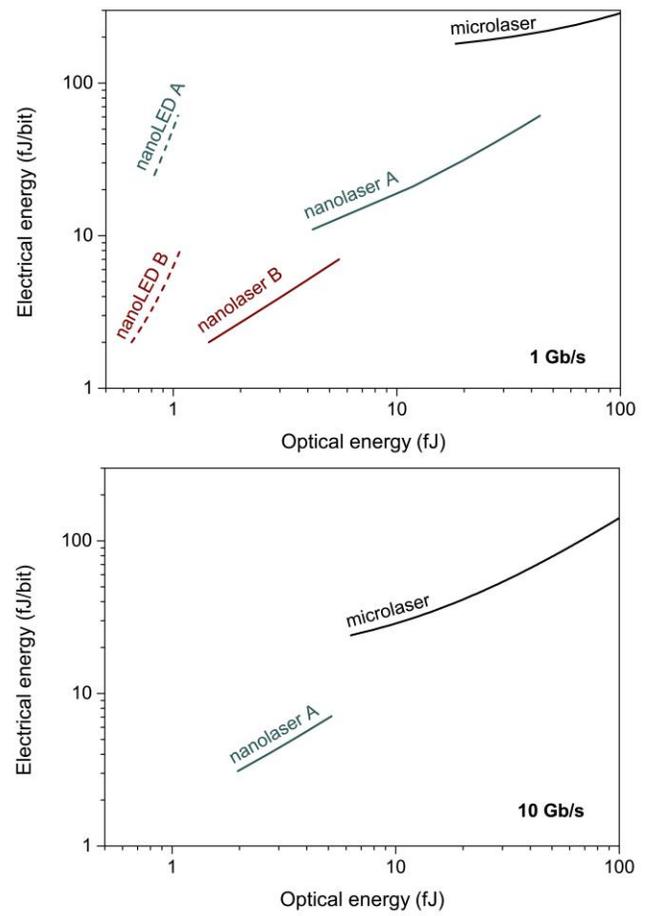

Fig. 8. Optical energy as a function of the electrical energy per bit for both laser (solid lines) LED (dashed lines) sources operating at data rates of (top) 1 Gb/s and (bottom) 10 Gb/s. The optical energy is calculated as $P \cdot T$, while for the electrical energy we assume $I \cdot V \cdot T$, taking $V$=1.0 V. The curves were plotted in the regions where the modulation bandwidths (see Fig. 5 and 7) are larger than the corresponding bit rate and in the regions where the current density is <100 kA/cm$^2$.

## V. Conclusion

In conclusion, we have analyzed some fundamental limits of scaling of nanoLEDs and nanolasers for optical interconnects. We have employed a physical model where the scaling of the spontaneous and stimulated emission rates with volume is consistently calculated for a semiconductor active region at room temperature, avoiding the use of *ad-hoc* parameters such as Purcell factor or spontaneous emission coupling factor. This has allowed us to derive the best-case scenario for (sub-)micrometer-sized optical sources operating at energies <10 fJ/bit. We have particularly looked at the possibility of producing 1-10 fJ of optical energy at current densities below 100 kA/cm$^2$. The main conclusions of our study are:

1) NanoLEDs with lateral dimensions >100 nm present much lower Purcell enhancements than theoretically possible for spectrally narrow emitters, which makes them unsuitable for direct modulation at rates >>1 Gb/s. Further downscaling would imply operation at unrealistically large current densities;

2) Nanolasers with active dimensions of few 100s nm have the potential to operate efficiently at low energy/bit levels and data rates up to 40 Gb/s. While their speed ideally increases with decreasing volume, practical limitations on current density will most probably favor devices with lateral dimensions of the order of the wavelength in the material (even not considering the inevitable optical loss in subwavelength cavities).

These conclusions should be viewed as an indication of the ultimate potential of small optical sources, rather than a prediction of performance of practical devices, which will be affected by effects not considered here, such as additional carrier losses, resistive voltage drop and heating. Also, while the absolute figures used for e.g. quality factor or maximum current density are somewhat arbitrary and may evolve with time, we claim that the general methodology employed here is an important tool to assess the potential of any proposal for ultrasmall light sources. Also, while we have focused our attention on the application in low-power optical interconnects, we expect that many of the considerations will also apply to other areas, such as sensing, as the optical power level impacts the resolution of a sensor.

## Acknowledgment

The authors would like to thank Victor Dolores-Calzadilla, Aura Higuera-Rodriguez and Meint Smit, Eindhoven University of Technology, and Dominik Heiss, Infineon Technologies, for fruitful discussions on metallo-dielectric nanoLEDs and nanolasers.

**Bruno Romeira** received the Ph.D. degree of physics (summa cum laude) and the title of European Ph.D. from the University of the Algarve, Portugal, jointly with the University of Glasgow, U.K., and the University of Seville, Spain, in 2012. He then held a Post-Doctoral Fellowship at the Microwave Photonics Research Laboratory, University of Ottawa, Canada, (2013–2014), and a Marie Skłodowska-Curie Research Fellowship at the Applied Physics Department and Institute for Photonic Integration of the Eindhoven University of Technology (2015–2017). He then joined the International Iberian Nanotechnology Laboratory (INL), Portugal, as a Marie Curie COFUND Research Fellow and he is since 2019 a staff researcher and coordinator of the EU-H2020-FET-OPEN project "ChipAI" at INL.

Dr. Romeira received the "Young Researchers Incentive Programme" Award from the Calouste Gulbenkian Foundation, Portugal, in 2009. He was a recipient of the "2011 IEEE Photonics Society Graduate Student Fellowship," from the IEEE Photonics Society, USA. He also received the "Best Ph.D. Thesis in optics and photonics in Portugal in 2012" from the Portuguese Society of Optics and Photonics. He is a member of Optical Society of America, USA.

**Andrea Fiore** received his PhD degree from the Univ. Orsay with a thesis on nonlinear optics. He then held postdoctoral appointments at University of California at Santa Barbara (1997-98) and at the Ecole Polytechnique Fédérale de Lausanne (1998-2001), and a researcher position at the Italian National Research Council (2001-2002). From 2002 to 2007, he led the activity on Quantum Devices at the Ecole Polytechnique Fédérale de Lausanne as assistant professor. Since October 2007, he holds the Chair of Nanophotonics at the Eindhoven University of Technology. Prof. Fiore has been the recipient of the "Professeur boursier" (Switzerland) and "Vici" (The Netherlands) personal grants, principal investigator in several national projects, coordinator of EU-FP6 project "SINPHONIA" and of Dutch national programs "Nanoscale Quantum Optics" and "Research Centre for Integrated Nanophotonics". He has been awarded the 2006 ISCS "Young Scientist" Award (International Symposium on Compound Semiconductors). He has coauthored over 150 journal articles and given more than 50 invited talks at international conferences.